# First order phase transition in a modified Ziff-Gulari-Barshad model with self-oscillating reactant coverages

I. Sinha*  and A. K. Mukherjee

*Department of Applied Chemistry,*
*Institute of Technology, Banaras Hindu University,*
*Varanasi 221005, India*
Fax: +91 542 2368428.
Email: isinha.apc@itbhu.ac.in

Abstract

Using kinetic Monte Carlo simulations, we study the effect of oscillatory kinetics due to surface reconstructions on Ziff-Gulari-Barshad (ZGB) model discontinuous phase transition. To investigate the transition, we do extensive finite size scaling analysis. It is found that the discontinuous transition still exists. On inclusion of desorption in the model, the order-parameter probability distribution broadens but remains bimodal. That is, the first-order phase transition becomes weaker with increase in desorption rate.

## 1. Introduction

Surface reaction models exhibit rich and complex variety of phenomena, including chaotic behavior, bistability, critical phenomena, out-of equilibrium phase transitions, etc. Understanding of such complex behavior can be helpful in explaining experimental results associated with catalysis and also for designing more efficient processes [1, 2]. The study of phase transitions in such non-equilibrium surface reaction models has attracted considerable attention. While a vast amount of research has been conducted on universality in continuous transitions [3, 4], comparatively little attention has been given to discontinuous transitions. The Ziff-Gulari-Barshad (ZGB) model [5], due to its simplified approach to modeling of catalytic oxidation of CO, gives us an opportunity to study the discontinuous transitions of a monomer-dimer surface reaction to a monomer-poisoned state. Many aspects of the ZGB discontinuous phase transition between the reactive and the poisoned states have been investigated [6]. These include propagation and fluctuation behavior of interfaces between active and poisoned states [5, 7, 8]; epidemic properties of an active droplet embedded in the poisoned state [9, 10]; and nucleation of droplets within the metastable active state [7, 11].

We note that more realistic models have been proposed for a better description of this technologically significant surface catalyzed reaction [12]. However, as mentioned above, the simple ZGB based models help us gain fundamental understanding of the effect of different factors on nonequilibrium phase transitions in surface catalyzed reaction systems. In fact, recently, a variant of the simpler stochastic lattice-gas Schloegl's second model for autocatalysis has been studied as a prototype of a nonequilibrium discontinuous phase transition [13-15].

The ZGB model assumes the catalytic surface to be a two-dimensional square lattice with periodic boundary conditions. This model follows the Langmuir–Hinshelwood mechanism for catalytic reaction [16].

$$CO(g) + S \rightarrow CO(S) \qquad (1)$$

$$O_2(g) + 2S \rightarrow 2O(S) \qquad (2)$$

$$CO(S) + O(S) \rightarrow CO_2(g) \qquad (3)$$

Here 'S' denotes a vacant site on the surface. The normalized probability $P_{CO}$ (proportional to the respective reactant pressure) determines the possibility that the next molecule to strike the surface is CO. Otherwise; an oxygen molecule may be selected to approach the surface with probability $1 - P_{CO} (= P_{O_2})$. The CO molecule from the gas phase can adsorb only onto a vacant site, while the oxygen molecule first dissociates into atoms and then may get adsorbed onto two neighboring vacant sites. The processes are irreversible once they occur. Adjacent O and CO react instantaneously to form $CO_2$, which desorbs leaving behind two empty sites. The Monte Carlo simulations, based on this picture of the surface, show two kinetic (or irreversible) phase transitions. Below a certain probability $(P_{CO} < P_1)$, the surface is oxygen poisoned, while above a higher CO partial pressure $(P_{CO} > P_2)$, the surface is CO poisoned. The first oxygen poisoning transition at $P_1$ is continuous, whereas the second transition (at $P_2$) to the CO poisoned state is discontinuous.

There are several aspects of this simplistic model that do not follow the experimental reality. Foremost among them is that real systems do not show the second-order transition to an oxygen-poisoned state [17-19]. On the other hand, discontinuous transitions between states of low and high CO coverage, governed by temperature changes, have been observed experimentally [20]. Above a certain critical temperature a smooth crossover replaces the transition. Inclusion of CO desorption rate $(k)$ in the ZGB model (popularly denoted as the ZGB-k model) was found to model this effect successfully. An earlier study by Tome et al. [11] estimated a critical desorption rate $k_c = 0.0406$ as the value at which the bimodal order-parameter probability distribution becomes unimodal. Preliminary and more extensive recent results on fourth order cumulant of CO coverage as well as finite size scaling analysis indicate the presence of a Ising model like critical desorption rate $(k_c)$ between $k = 0.03$ and $0.04$ [11, 21]. Above $k_c$ the transition becomes a smooth crossover. The inclusion of the Eley-Rideal (ER) step in the ZGB model results in termination of the coexistence curve at a value of $k_c$ that decreases with the probability of the ER step [22]. Subjecting the model to symmetrically oscillating reactant pressure leads to the smoothening of the ZGB-k phase transition [23]. The introduction of an asymmetrically oscillating reactant pressure makes the transition second-order. Fourth order cumulants and finite size scaling evidence suggests that this far from equilibrium phase transition belongs to the equilibrium Ising universality class [24].

An extensively studied aspect of catalytic oxidation of CO on Pt-group metal surfaces is the oscillatory kinetics observed both in the low pressure and the high reactant pressure conditions. Oscillations in the low-pressure limit are due to coupling of surface reconstructions with the reactant coverage [25]. In contrast, oscillations at higher pressures (up to atmospheric) may arise because of coupling of the CO oxidation reactions with surface oxidation [25-27]. Attempts to develop a theoretical understanding of this problem include analytical mean-field and pair approximation approaches [28-30]. These use variables representing the surface structure e.g., hexagonal versus square coupled with variables representing the surface concentrations. On the other hand, Monte Carlo simulations and lattice gas cellular automata address the mesoscopic mechanisms producing these oscillations and therefore have been preferred in recent publications.

It is known from experiment that the surface reconstruction takes place if the coverage of certain adsorbates (e.g., CO or NO) increases above or drops below a certain critical value [25]. Based on this, within the ZGB framework, Albano proposed a modified model, which demonstrated oscillatory behavior of coverage due to surface reconstructions [31, 32]. In this study an arbitrary CO surface coverage $(\theta_{CO})$ fraction was used to define surface reconstruction. That is, when $\theta_{CO} < 0.1$ the surface reconstructs globally and only CO (but no $O_2$) can adsorb on to the reconstructed phase. On reaching $\theta_{CO} = 0.485$, all surface sites change to the unreconstructed phase which allows $O_2$ adsorption. The proposed model demonstrated regular oscillations in $CO_2$ reaction rate. However, such regular oscillations have been experimentally observed only in the Pt(100) system and that also for small number of periods [33, 34]. To explain different facets of experimental observations, models based on the assumption that the formation of patches of CO surface coverage could lead to local surface reconstruction have also been proposed [32]. Kortluke et al. [35] further extended the local reconstruction approach and proposed that individual sites on the surface could also transform to the reconstructed phase. More recently, Provata et al. [36] have again used a modified global reconstruction model. Unlike earlier models in which the restructured phase was assumed on the basis of its adsorbate preference, here direct surface lattice reconstruction was achieved. On assumption of global surface reconstruction for high lattice coverage of one of the reactive species, regular $CO_2$ reaction rate oscillations similar to that in Albano's model were observed.

For both global and local reconstruction models, on taking the long time average of the oscillating CO coverage for a given $P_{CO}$ as the order parameter, a phase diagram showing different regimes of reactivity with respect to the applied $P_{CO}$ is obtained. The introduction of reconstruction naturally leads to the disappearance of the continuous transition to oxygen poisoned state present in the classical ZGB model. Further, the phase diagram obtained by assuming global reconstruction also suggests (similar to the classical ZGB model) that there is a discontinuous transition from the oscillatory reactive to a completely poisoned state at a critical value of $P_{CO}$ [31]. Alternatively, the assumption of local reconstruction leads to the rounding of this transition [37]. The present study focuses on the suggested discontinuous phase transition in the global reconstruction model [31]. As is evident from the foregoing discussion, this model does not take into account realistic experimental conditions. However, such a simplistic study does give us an opportunity to study the effect of oscillatory kinetics on a discontinuous transition. We, therefore, first do an extensive statistical analysis on this system analogous to those carried out in recent literature [11, 22-24] on non-equilibrium ZGB based model systems. The analysis confirms that indeed there exists a first order irreversible transition in the phase diagram of this 'global surface reconstruction' model [31]. Then the effect of CO desorption on this phase transition is investigated. We compare our results with the extensive results available on phase diagram of the earlier mentioned ZGB-k model. Thus, at increasing desorption rates we compute the order-parameter probability distribution in the transition region. An extensive finite size scaling analysis of the transition is presented.

## 2. Model and simulation procedure

Under low-pressure regimes, the catalytic oxidation of CO on Pt surfaces exhibits oscillatory behavior due to the difference in the sticking coefficients for $O_2$ on the different surface phases. The transition between the surface phases is adsorbate induced. Thus, the adsorption of CO up to a certain threshold leads to a first order transition of the reconstructed surface phase [hex on Pt (100)] to the non-reconstructed surface phase [1x1 on Pt (100)]. In this study, we consider the global reconstruction model of Albano approximating the Pt (100) case. The ratio of the $O_2$ sticking coefficients on Pt (100) between reconstructed and non-reconstructed is $<10^{-2}$. Keeping in view the wide difference between the $O_2$ sticking coefficients of the reconstructed and non-reconstructed phases, their corresponding model values are assumed to be 0 and 1, respectively. In context of the present model, once an oxygen molecule has been selected to approach the catalyst surface (with probability $1-P_{CO}$), the sticking coefficient is the probability of adsorption of the dissociated oxygen atoms on to adjacent vacant sites.

The simulations start from an empty square lattice of side length $L$. A time unit or the Monte Carlo step (MCS) of our simulation involves $L^2$ trials, i.e. during a time unit each site of the lattice is visited once on an average. Each simulation, at a fixed value of $P_{CO}$, consists of the following sequence of events. The initial state of the surface corresponds to the reconstructed phase with the oxygen-sticking coefficient equal to zero. Effectively, this means that only adsorption of CO occurs in a random manner on the vacant sites in model catalyst surface at a rate proportional to $P_{CO}$. Once CO coverage $(\theta_{CO})$ exceeds $\geq 0.485$, all sites in the model catalyst surface transform to the non-reconstructed phase. The sticking coefficient of oxygen becomes 1, the normal ZGB algorithm is initiated and $\theta_{CO}$ decreases because of reactions between neighboring adsorbed CO and O. Whenever the condition $\theta_{CO} \leq 0.1$ is achieved, all sites in the system are assumed to transform to the reconstructed phase. Thereafter, as mentioned earlier, only CO adsorption occurs at a rate proportional to $P_{CO}$ until the condition $\theta_{CO} \geq 0.485$ is achieved again. This completes one cycle of the adsorbate coverage induced transition. The critical adsorbate coverage conditions ($\theta_{CO} \geq 0.485$ and $\theta_{CO} \leq 0.1$) at which the catalyst surface undergoes phase transitions, and the corresponding oxygen sticking coefficients are typically the values used by Albano in his model [28]. The purpose of using these values is to enable comparison with the results of reference [28]. It must be mentioned that the values used by Albano are arbitrary. Changing the critical adsorbate coverage conditions ($\theta_{CO} \geq 0.485$ and $\theta_{CO} \leq 0.1$), as described above, affects the results only qualitatively [28].

The simulations are then carried out at different desorption rates $(k)$, ranging from 0 to 0.04. In this case, for each reactant pressure, a Monte Carlo simulation is carried out by generating a sequence of adsorption (with probability $1-k$) and desorption (with probability $k$) trials. Rest of the algorithm remains the same.

## 3. Results and discussion

Corresponding to $\theta_{CO}$, the oxygen coverage and $CO_2$ production rate are denoted by $\theta_O$ and $R_{CO_2}$ respectively. Fig. 1a, 1b and 1c show the oscillatory behavior of $\theta_{CO}$, $\theta_O$ and $R_{CO_2}$ respectively with increase in simulation time (in MCS) at $k=0$. The results are presented at $P_{CO} = 0.5238$, which is near the transition, on the reactive side. There is some fluctuation in both the period and amplitude of the oscillations. This is similar to the results given in reference [28] for near transition regimes. However, over longer periods of time, the overall oscillatory behavior seems to be in a steady range. All such simulations at and near the transition are carried out for $10^5$ MCS time length. Out of these, the initial $7 \times 10^4$ steps are disregarded. Keeping in view the oscillatory nature of $R_{CO_2}$ and $\theta_{CO}$, we take their time average over the next $3 \times 10^4$ steps and denote them as $\overline{R}_{CO_2}$ and $\overline{\theta}_{CO}$. The latter $\overline{\theta}_{CO}$ is assumed to be the order parameter. Equivalent results are obtained even if $\bar{R}_{CO_2}$ is considered to be the order parameter. It is important to mention that recent phase transition studies on modified ZGB model systems subjected to oscillating reactant pressure have also assumed the time averaged $\theta_{CO}$ or time average $R_{CO_2}$ to be the order parameter [23, 24].

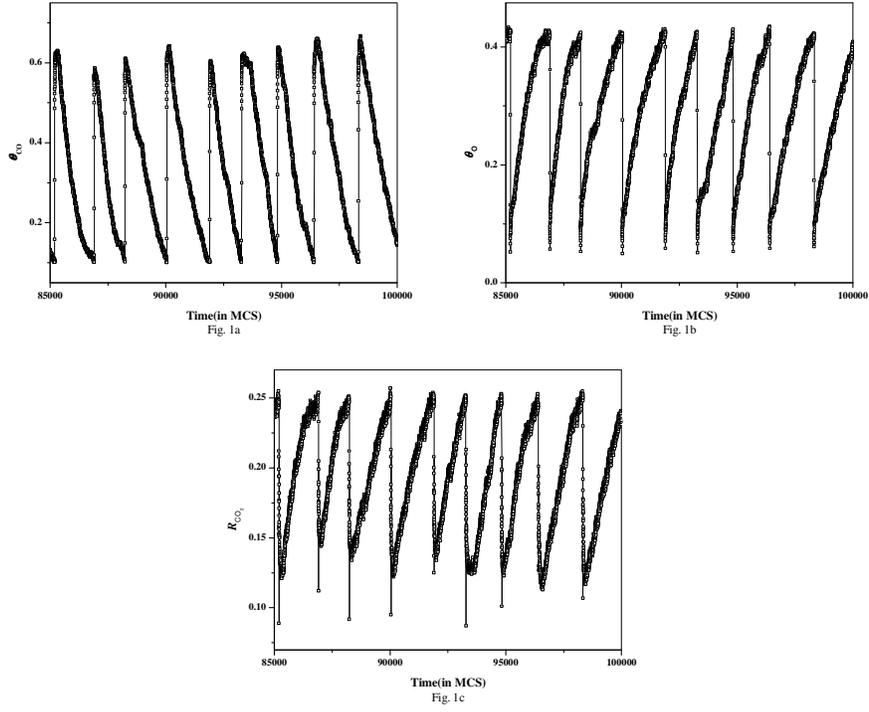

Fig. 1 Oscillations in system parameters with time (in MCS). a) $\theta_{CO}$ versus time, b) $\theta_{O}$ versus time and c) $R_{CO_2}$ versus time plots.

The probability distribution for $\bar{\theta}_{CO}$ (denoted by $P(\bar{\theta}_{CO})$) is evaluated using the following procedure. The number of times $(N_i)$ the order parameter falls in the interval $[(i-1)\Delta, (i)\Delta]$ is recorded. Here the magnitude of the interval $i$ is $\Delta = 0.01$ following $1 \leq i \leq 100$. The total number of Monte Carlo simulations is given by $\sum_i N_i = N$. The number of simulations $(N)$ required for generating sufficient statistics is computed by $N \times (L \times L) = 4.5 \times 10^5$, for instance, $N = 1250$ when $L = 60$. The probability that $\bar{\theta}_{CO}$ falls in the interval $i$ is given by $p_i = N_i/N$. Figure 2 presents the probability distribution for $\bar{\theta}_{CO}$ against $P_{CO}$ at $k = 0$ for two different system sizes ($L = 180$ and $120$) near the coexistence region. The figures show the narrow range of $P_{CO}$ values where the probability distribution is bimodal, indicating a first order transition. Away from this region the distribution becomes unimodal. At the coexistence point, the areas under both peaks are equal [38, 39]. However, the reactive side distribution in the figures show broadening, making difficult a direct determination of the finite size coexistence value by this method.

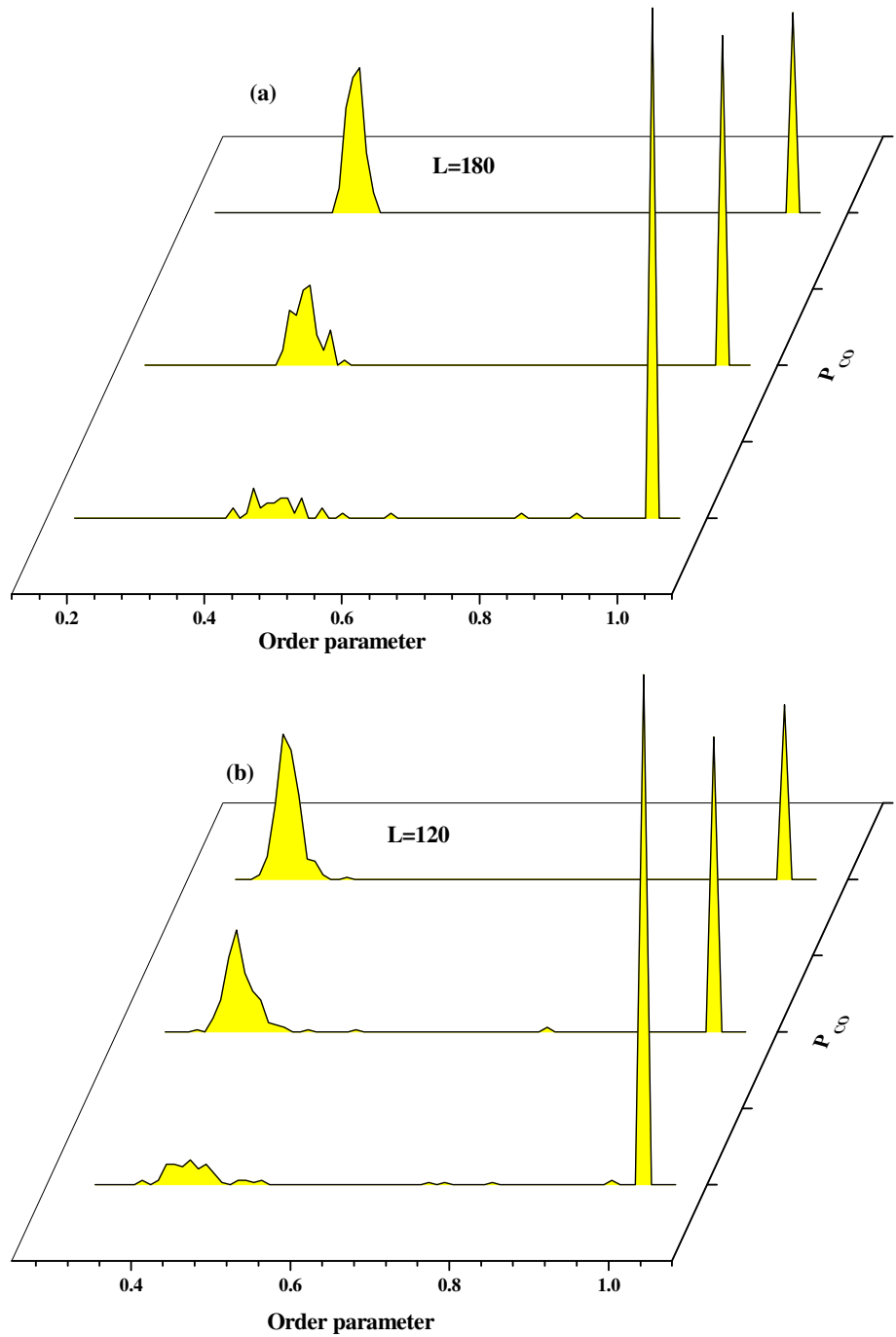

Fig. 2 Order-parameter $(\bar{\theta}_{CO})$ probability distribution at different $P_{CO}$ values showing the narrow coexistence region at $k=0$ at a) $L$=180 and b) $L$=120.

To confirm that there indeed exists a discontinuous transition between the reactive and poisoned state for $k=0$, we employ the finite-size scaling (FSS) method established in recent publications for analyzing such nonequilibrium surface phase transitions [21-24, 11]. The finite size scaling analysis of the fluctuations in $\bar{\theta}_{CO}$ in a $L \times L$ system was evaluated by the following expression.

$$\chi_L = L^2 \left( \left\langle \bar{\theta}_{CO}^2 \right\rangle_L - \left\langle \bar{\theta}_{CO} \right\rangle_L^2 \right) \qquad (4)$$

Fig. 3a depicts $\chi_L$ versus $P_{CO}$ plots for systems ($k=0$) with different sizes. As expected, the maximum values of the order-parameter fluctuation curve shift and increase in height with increasing $L$. That is, the peak positions of $\chi_L$ approach the infinite transition point with increasing $L$. For an equilibrium first-order phase transition the relation $\chi_L^{max} \sim L^d$ should be followed. Here $d$ is the spatial dimension of the system. In Fig. 3b we plot $\ln\left(\chi_L^{max}\right)$ against $\ln(L)$. A perfectly linear fit is obtained with a slope 1.99. This agrees well with the expected value of $d=2$ for a first order transition [22, 23].

Several recent studies have demonstrated that the fourth-order reduced cumulant of the order parameter is an effective approach to locate and classify even non-equilibrium phase transitions [11, 21-24]. This is a tool which quantifies the shape of the order-parameter distribution. The fourth-order cumulant, taking $\bar{\theta}_{CO}$ as the order parameter [11] is given by the following expression.

$$u_L = 1 - \frac{\mu_4}{3\mu_2^2} \qquad (5)$$

where $\mu_n = \left\langle \left(\bar{\theta}_{CO} - \left\langle \bar{\theta}_{CO} \right\rangle\right)^n \right\rangle_L = \int_0^1 \left(\bar{\theta}_{CO} - \left\langle \bar{\theta}_{CO} \right\rangle\right)^n P(\bar{\theta}_{CO}) d\bar{\theta}_{CO} \qquad (6)$

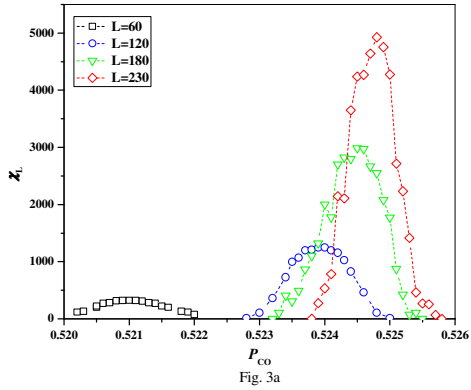
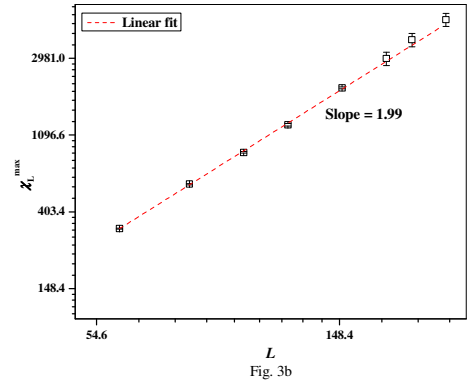

Fig. 3 a) The order-parameter fluctuation $\chi_L$ against $L$ plots for different system sizes $(k=0)$. The dotted lines joining the data points are guides to the eye. b) Plot of $\chi_L^{max}$ versus $L$ on ln scale. $\chi_L^{max}$ is the maximum value of $\chi_L$ taken from part a) of the figure.

denotes the $n$ th central moment of $\bar{\theta}_{CO}$ and $P(\bar{\theta}_{CO})$ is the probability distribution for $\bar{\theta}_{CO}$. The equal area bimodal distribution corresponding to coexistence yields a positive maximum for the cumulant versus $P_{CO}$, flanked on either side by negative minima and approaching zero far away from the transition. In Fig. 4, we show the variation of $u_L$ with $P_{CO}$ for different system sizes with desorption rate $k=0$. Plots for all system sizes display the typically expected behavior in the coexistence region. That is, a positive maximum of the cumulant with adjacent negative deep minima and tending to zero as one moves away from the transition. The negative deep minima on either side correspond to the beginning and the end of the coexistence region. The maximum of $u_L$ defines the $L$-dependent transition point. For systems with $k=0$, at all sizes considered, the maximum values of the cumulants $u_L^{max}$ are at or tend to the value $0.66 \sim 2/3$ (Fig. 4a and 4b). Figure 4b gives a magnified view of the regions around $u_L^{max}$. These results are in agreement with the

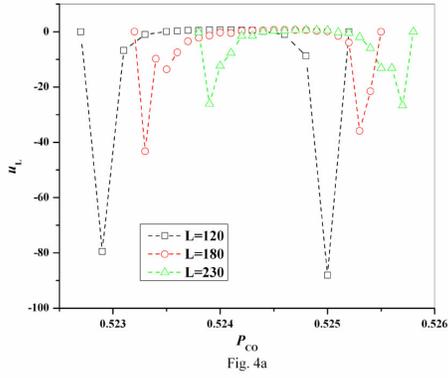
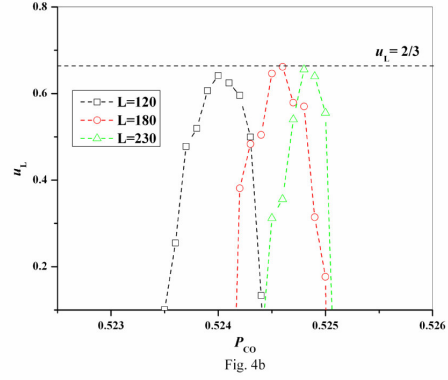
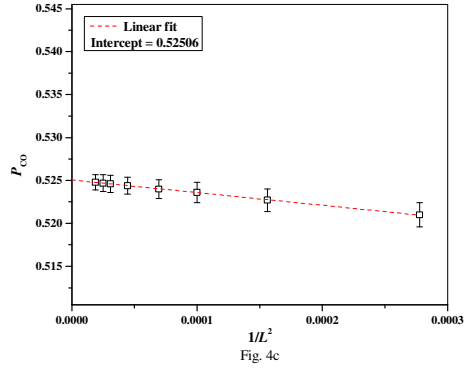

Fig. 4 a) The fourth-order cumulant $u_L$ versus $P_{CO}$ for systems of different sizes $(k=0)$. b) Magnified view of the same plot showing region near $u_L^{max}$. The lines joining the data points are guides to the eye. The errors are of the order of 1% for the highest values of $u_L$. c) The finite size coexistence $P_{CO}^c(L)$ value versus $1/L^2$ for systems at $k=0$. The intercept of the weighted least-squares linear fit yields the estimated $P_{CO}^c$ at $L \to \infty$.

earlier studies by Tome et al. [21] and recently Machado et al. [11] on the ZGB-k model for $k < k_c$. These authors reported that analogous to the equilibrium Ising model [40] results, $u_L \to 2/3$ for $k < k_c$ and $u_L \to 0$ for $k > k_c$. It must be mentioned that for all system sizes the $P_{CO}$ values at which $u_L^{max}$ and $\chi_L^{max}$ values occur are the same, thus uniquely identifying the finite size first-order transition or coexistence value.

The finite-size scaling theory of equilibrium first-order phase transitions implies that the shift in the position of the transition in a finite system of linear size $L$ with periodic boundary conditions is inversely proportional to the system volume, $L^d$ [41, 42] (here the dimension $d=2$). Although there is no analogous scaling theory for the non-equilibrium systems, however, phase transition studies on ZGB based models have successfully used the same scaling relation [11] as for equilibrium cases.

$$P_{CO}^c(k,L) - P_{CO}^c(k) \propto L^{-2} \qquad (7)$$

Here $P_{CO}^c(k)$ is the transition value of the CO adsorption rate $P_{CO}$ in the infinite-$L$ limit. In Fig. 4c, we plot the finite size $P_{CO}^c(L)$ for $k=0$ versus $L^{-2}$ for $L=$ 60, 80, 100, 120, 150, 180, 200 and 230. As one observes the points fit a straight line very well. The correlation coefficient of the linear least squares fit is >0.999. The intercept gives the value of $P_{CO}^c(k=0)$ as 0.5250(6). This value is close to the first-order transition point 0.5256 of the regular ZGB model [5]. We note that Albano [28] had suggested the value $P_{CO} = 0.5235 \pm 0.0005$ as the transition point.

To understand the effect of desorption on this discontinuous transition, we now present results for systems with $k>0$. Figures 5a and 5b show the order parameter probability distribution for systems with sizes $L=60$ and 180 respectively at different desorption rates. In each figure order-parameter distributions are shown for three different $P_{CO}$ values in narrow transition region. The $P_{CO}$ value in the middle is the finite size coexistence or transition values (denoted as $P_{CO}^c(L)$) value at which $u_L^{max}$ and $\chi_L^{max}$ values are observed for the system. For systems with $L=60$ and $k=0.01$, besides the clear bimodal distribution, there is also broadening between the reactive and the partially poisoned sides. This broadening characteristic of the distribution increases significantly for $k = 0.02$ and more so for $k = 0.03$. Finally, for $k = 0.04$, a broadened distribution peak moves from the reactive side to the partially poisoned side as $P_{CO}$ is increased (Fig. 5a). That is a unimodal distribution sweeps across smoothly from the reactive side to the partially poisoned side. In the Fig. 5b we consider the effect of desorption rate on much larger systems ($L = 180$). Here, in contrast to the observations made for $L = 60$, the broadening of the order parameter distribution is much lesser. However, the peak broadening still occurs and becomes significant for $k \geq 0.03$. At the $(L = 180)$ transition $P_{CO}^c$ value, despite the broadening we are able to approximately make out predominantly bimodal nature of the distribution. Here, even at $k = 0.04$, it is a broadened bimodal type distribution. Clearly the nature of the transition changes with the system size.

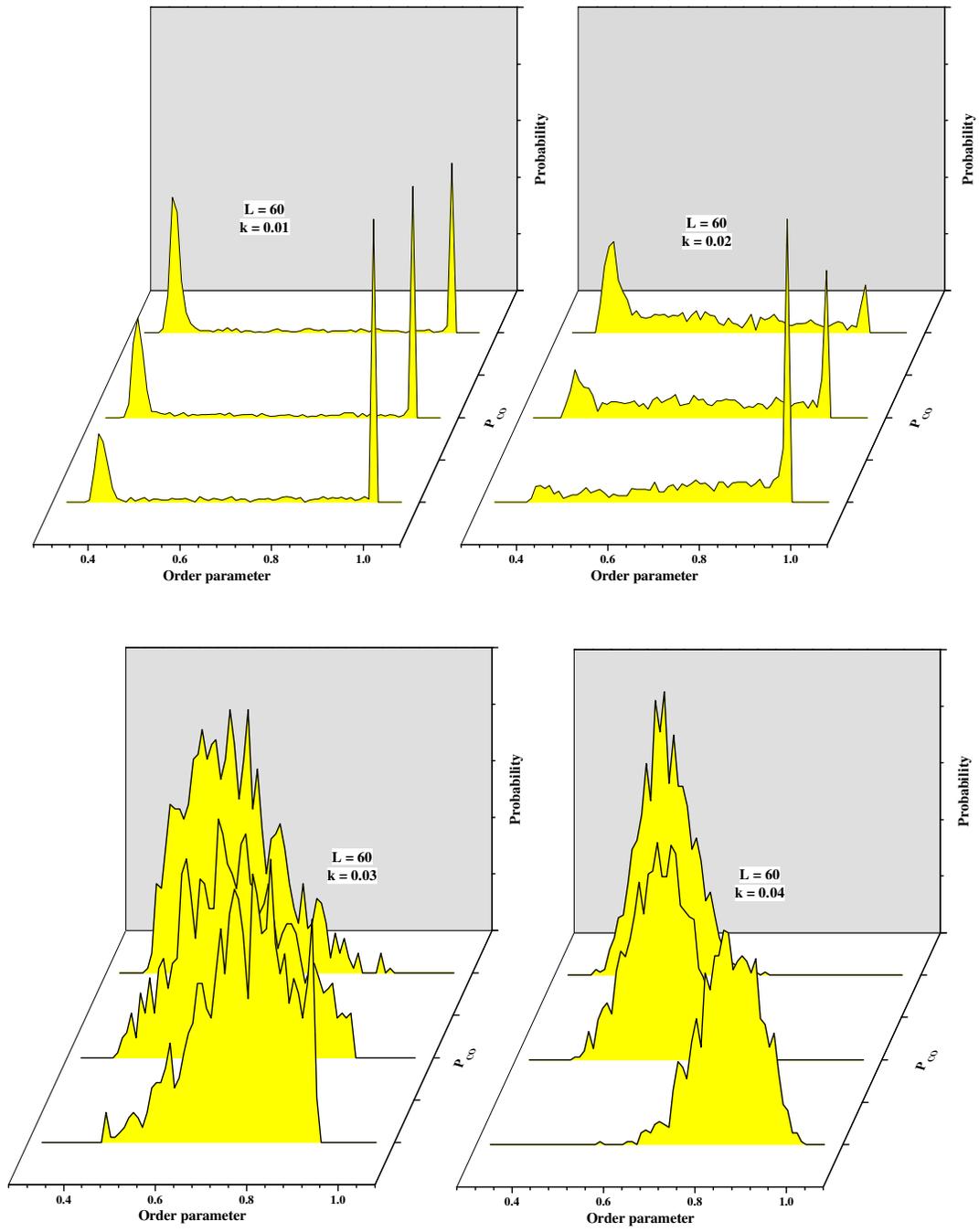

Fig. 5a

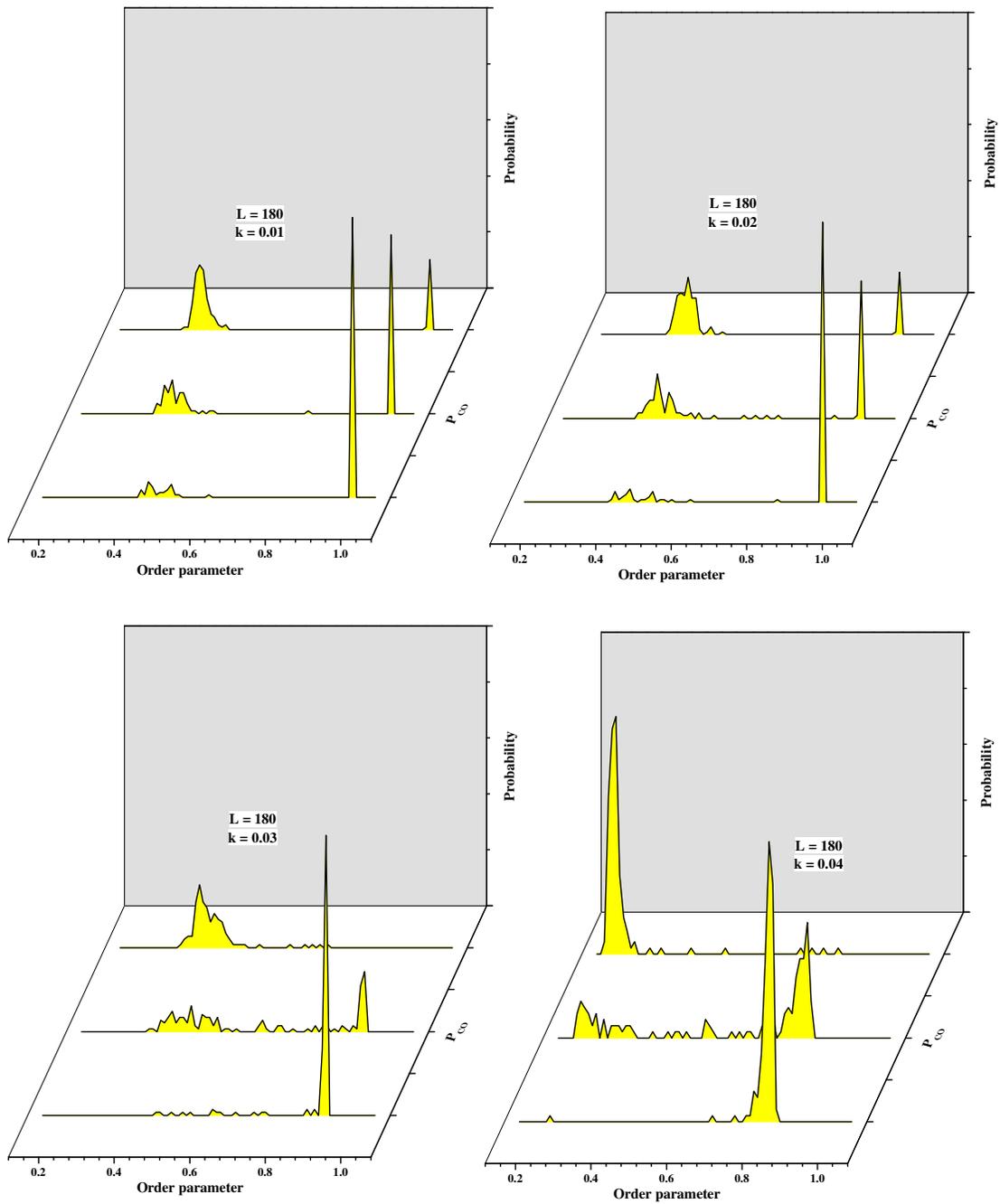

**Fig. 5b**

Fig. 5 Order parameter probability distributions $\left(\bar{\theta}_{CO}\right)$ at $P_{CO}$ values in the narrow coexistence region for system of a given size at different desorption rates $(k > 0)$. a) $L = 60$ and b) $L = 180$.

Figures 6a and 6b now focus only on the distribution at the finite size coexistence point $P_{CO}^c(L)$ for systems at different desorption rates. Figures are given for two system sizes. For both system sizes, we observe increase in desorption rate make probability distribution peaks more broadened. Thus, in Fig. 6a $(L = 120)$ at $k = 0.04$ the bimodal distribution becomes very much broadened making it difficult to distinguish between the two peaks. On considering a given large enough system size ($L = 230$), we find the transition retains the bimodal distribution character (Fig. 6b) even at high desorption rates ($k = 0.04$). Till $k = 0.03$, significant broadening is only manifested for the reactive side order-parameter peak. For $k = 0.04$ the partially poisoned side order-parameter peak broadening also becomes significant. We conclude that for larger system sizes, the order-parameter peak broadening increases with desorption rate, while retaining the essential bimodal character. This suggests that the first-order transition becomes weaker with increase in desorption rate. Fig. 6c describes the effect of system size on the order parameter probability distribution at $k = 0.03$. As the system size increases the bimodal nature of the distribution is more clearly manifested.

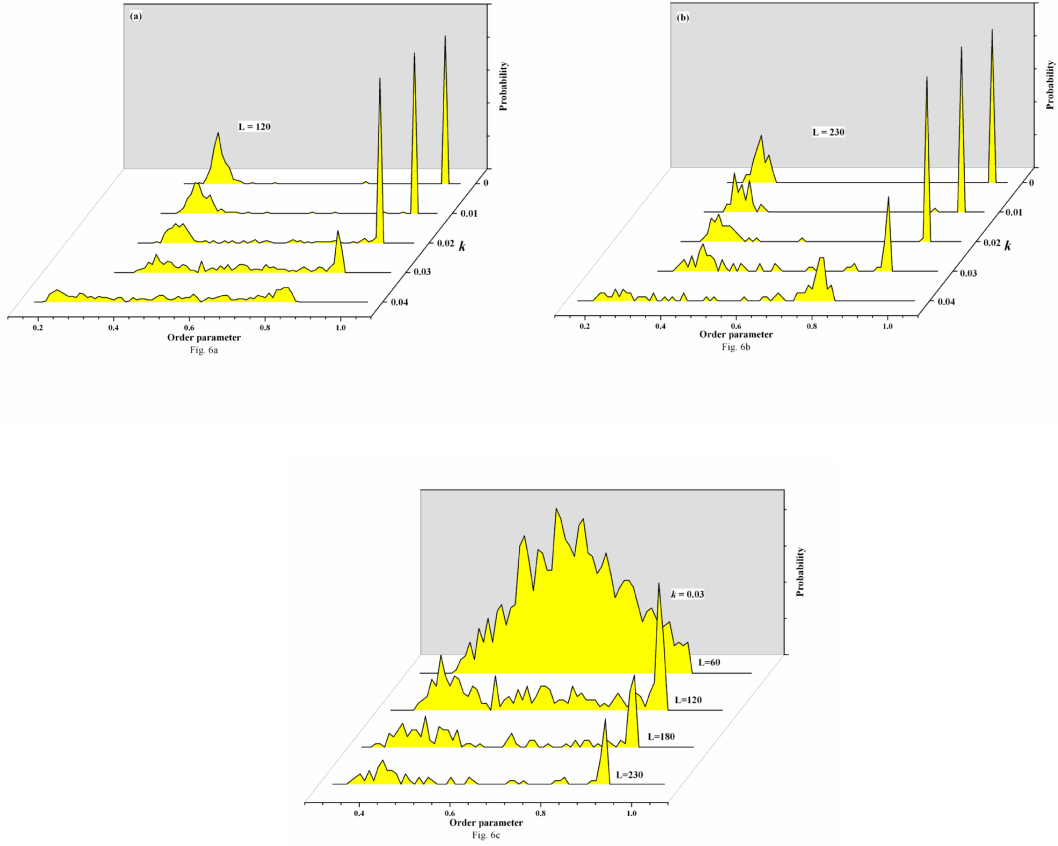

Fig. 6 A comparison of order-parameter $\left(\bar{\theta}_{CO}\right)$ probability distributions for systems subjected to increasing desorption rates. a) $L = 120$ and b) $L = 230$. The distributions are for the finite size coexistence $P_{CO}^c$ value at which $u_L^{max}$ and $\chi_L^{max}$ are obtained. c) order-parameter $\left(\bar{\theta}_{CO}\right)$ probability distributions at $P_{CO}^c$ for different system sizes at $k = 0.03$.

To have a more quantitative perspective of the effect of system size on the phase transition we perform extensive finite size scaling analysis over sizes $L$ = 60, 80, 100, 120, 150, 180, 200, 230. Figures 7a and 7b show $\ln\left(\chi_L^{max}\right)$ against $\ln(L)$ plots at different

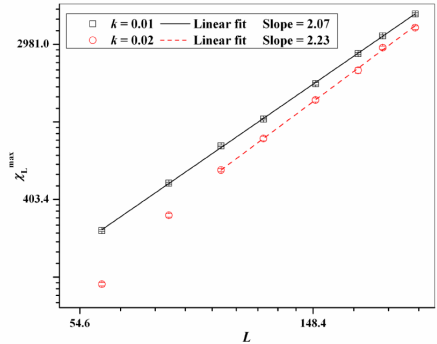 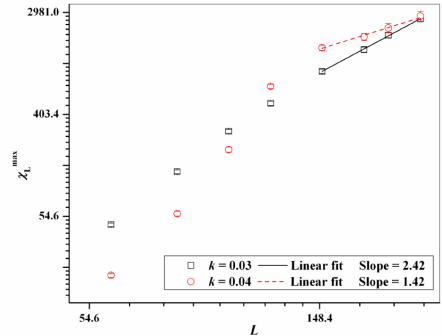

Fig. 7a  Fig. 7b

Fig. 7 Plots of $\chi_L^{max}$ versus $L$ on ln scale, a) for systems with $k = 0.01$ and $0.02$, b) for systems with $k = 0.03$ and $0.04$.

desorption rates. For $k = 0.01$, the slope obtained is 2.07, which decreases to a value closer to 2 when we exclude the $L = 60$ data point from the linear fit. On further increasing $k$ to 0.02, a good linear fit is obtained only after excluding the sizes $L = 60$ and 80. And when $k = 0.03$, the linear fit shifts to only the last four data points, that is $L > 120$. As seen from the figures 7a and 7b, the slopes obtained even with this limited set of data points greatly exceed the first order transition requirement $d = 2$. However, for $k = 0.04$ the slope for the linear fit over data points (for systems with $L > 120$) finally decreases to a value less than 2. Therefore, with increase in desorption rates, the range of system sizes over which linear fit can be obtained shifts to larger sizes and also such linear fits have anomalous slope values. For a given desorption rate, evidently the transition behavior changes with increase in system sizes, which is in agreement with Fig. 6c. With desorption rate, the deviation from the ideally expected $d = 2$ value for a first-order transition also increases. As mentioned earlier, the broadening of the order-parameter probability distribution occurs for $k = 0.04$ even for larger system sizes and this seems to be the reason for decrease in slope of the respective

$\ln\left(\chi_L^{max}\right)$ against $\ln(L)$ plot to a value much lesser than the expected one. The finite size scaling of the systems with $k > 0$ reinforces our earlier inference that the first-order transition becomes weaker with increase in desorption rate.

Now in a typical Ising universality class model like the ZGB-k system one observes the bimodal distribution at $k < k_c$ becoming unimodal at $k > k_c$. The boundary between the bimodal to unimodal distribution defines $k_c(L)$ [11, 21]. As is clear from the observations, in the present study we fail to observe any such change over. The transition remains first-order and only weakens at higher desorption rates. The present model is therefore unlike the ZGB-k model and does not belong to the Ising universality class.

It must be mentioned that in the reference [21], the critical point $k_c$ in the ZGB-k model was estimated by both the histogram method as well as by value of the fourth order cumulant. The $k_c$ values thus obtained were found to be in agreement. These authors found that $u_L^{max} = 0.61$ at $k_c$ as expected for the two dimensional Ising universality class. Machado et al. [11] later repeated the cumulant studies for larger system sizes and also did extensive finite size scaling analysis to confirm the results in reference [21]. As has been already mentioned, in the present case there is only first-order transition which gets weaker with desorption. The second-order critical point, if it exists, may be located from the common intersection point of different system size $u_L^{max}$ against $k$ plots. Figure 8 shows $u_L^{max}$ against $k$ plots with respect to different system sizes for the present model with oscillatory kinetics due to surface reconstruction. We observe (Fig. 8) there is no common intersection point of the plots of different sizes. The decrease in $u_L^{max}$ values with $k$ is therefore due to the broadening of peaks in the bimodal order-parameter probability distribution at the coexistence value.

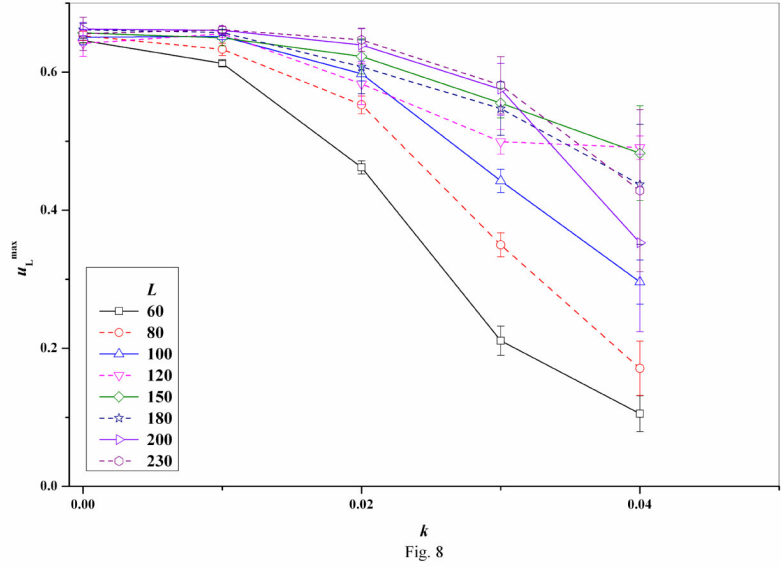

Fig. 8

Fig. 8 Fourth order cumulant $u_L^{max}$ versus $k$ plots for different system sizes.

Here it is pertinent to mention that recent investigations have revealed the fully frustrated cubic lattice Ising spin model also displays a weak first-order phase transition [43]. These authors also used the histogram technique to arrive at this result. In the present model, the stronger coupling of the local reaction dynamics with the global surface reconstruction with increase in desorption rate may be the reason for the weakening of the first-order transition behavior.

## 4. Conclusions

A modified ZGB model including oscillatory kinetics due to adsorbate-induced reversible transitions in the structure of the catalyst has been considered. Extensive kinetic Monte Carlo simulations have been carried out to study the phase transition in this model. The modified ZGB model shows the first-order phase transition when there is no desorption. This has been concluded from finite size scaling numerical analysis. On increasing the desorption rate (for larger system sizes) the order-parameter probability distribution broadens but the bimodal character of the distribution is preserved. This suggests that the phase transition remains first-order and only becomes weaker with increase in desorption rates. For a given desorption rate the strength of first-order phase transition increases with size. Due to this the finite size scaling analysis gives anomalous results for systems at $k > 0.01$.

Acknowledgements

The authors wish to thank Professor Per Arne Rikvold (Florida State University, USA) for his helpful suggestions. An extensive part of the computations was carried out at the central computing facility of BHU.